# New generalized uncertainty principle with parameter adaptability for the minimum length

Xin-Dong Du[*], Chao-Yun Long[†]

Department of Physics, Guizhou University, Guiyang 550025, China

**Abstract**

There have been many papers suggesting that the parameter of the generalized uncertainty principle should be negative rather than positive in some specific scenarios, and the negative parameter can remove the minimum length. However, the minimum length is a model-independent feature of quantum gravity and it should not be affected by the specific scenarios. In order to solve this contradiction, we derive a new generalized uncertainty principle to reflect a fixed and unified minimum length in both cases of positive and negative parameters.

## 1. Introduce

Various theoretical models of gravity and quantum mechanics have predicted the existence of a minimum length [1-3]. However, the Heisenberg uncertainty principle indicates that the minimum observable length is actually zero. Thus, the Heisenberg uncertainty principle should be modified to reflect the existence of a minimum length, and the corresponding corrected result is known as the generalized uncertainty principle [4-6], namely:

$$\Delta x \Delta p \geq \frac{\hbar}{2}\left(1 + \frac{\alpha L_p^2}{\hbar^2}(\Delta p)^2\right), \tag{1}$$

where $\Delta x$ and $\Delta p$ are the uncertainties of position and momentum, $L_p = \sqrt{G\hbar/c^3}$ is the Planck length, $\alpha$ is a dimensionless parameter and the above inequality can return to the Heisenberg uncertainty principle as $\alpha = 0$. Although the parameter $\alpha$ is not explicitly defined, it is often regarded as a positive value to keep the minimum length $(\Delta x)_{min} = \sqrt{\alpha} L_p$ existing.

In recent years, there have been many papers suggesting the parameter of the generalized uncertainty principle should be negative rather than positive in some specific scenarios such as: a metric to reproduce the Hawking temperature deformed by the generalized uncertainty principle [7], an uncertainty relation from a crystal-like universe [8], the sparsity of Hawking radiation affected by rainbow gravity [9], the Magueijo and Smolin formulation of doubly special relativity [10], and a method to restore the Chandrasekhar limit modified by the generalized uncertainty principle [11, 12]. Therefore, different quantum gravitational models might lead to different parameter signs (positive or negative), and the parameter sign should be treated as a changeable object. However, an inevitable contradiction will arise from the above conclusion: on the one hand, the existence of a minimum length is a common fact that should not be affected by the specific scenarios [13, 14]. On the other hand, when the parameter is negative, the minimum length will be removed from the generalized uncertainty principle and physics can become classical again [15, 16]. In other words, the generalized uncertainty principle does not adapt to the case of negative parameter. The purpose of this paper is to modify the generalized uncertainty principle to let it reflect a fixed and unified minimum length in both cases of positive and negative parameters.

---

[*] First author.
[†] Corresponding author. E-mail: LongCYGuiZhou@163.com

Usually the generalized uncertainty principle can be obtained from the commutation relation that owns a more general form [17], so the ways to revise the generalized uncertainty principle often depend on modifying commutation relations. For example, in [18], a modified commutation relation is proposed to lead to a new generalized uncertainty principle to reflect the maximum momentum, agreeing with doubly special relativity. But to strictly change the structure of the commutation relation is difficult, and almost papers avoid this point to give their assumptions for improved commutation relations rather than related derivations [17-21]. Besides, using modified commutation relations seems to violate the equivalence principle [22], so there are various challenges in revising the generalized uncertainty principle from modified commutation relations. In fact, it is entirely possible to revise the generalized uncertainty principle without modifying commutation relations. In this paper, we are going to show that the generalized uncertainty principle can be modified based on a black hole entropy influenced by the back reaction [23], and the corresponding corrected generalized uncertainty principle will succeed in reflecting a fixed and unified minimum length under different parameter signs.

This paper is organized as follows: in Sect. 2, to guarantee the feasibility and effectiveness of the follow-up correction process, we clarify an equivalence relation between the generalized uncertainty principle and the black hole entropy. In Sect. 3, an entropy corrected by the back reaction is used to derive a new generalized uncertainty principle. In Sect. 4, we will see that the new generalized uncertainty principle can maintain the existence of a fixed and unified minimum length regardless of whether its parameter is positive or negative. In Sect. 5, some other generalized uncertainty principles are given to be compared with the new one. In Sect. 6, we discuss the equivalence relation and the application range of the new generalized uncertainty principle.

## 2. Generalized uncertainty principle and entropy

It is well known that the black hole entropy can be corrected by the generalized uncertainty principle [24]. Therefore, if the generalized uncertainty principle could be reproduced from the corrected entropy, then the equivalence relation between them would be regarded as valid and provide us with an effective path to indirectly modify the generalized uncertainty principle by improving the entropy. In contrast, without the equivalence relation, some unpredictable consequences would be obtained from the improved entropy. For example, the modified generalized uncertainty principle might even fail to reflect the minimum length in any situation. To avoid these uncontrollable modification results, it is necessary for us to first find an existent equivalence relation. That is to say, the equivalence relation is the theoretical guarantee of our subsequent correction process.

However, a general equivalence relation is too hard to be built up, because there are all kinds of black holes so that it would be difficult to express their corresponding entropies in a unified form. In this section, we are going to construct an equivalence relation between the generalized uncertainty principle in Eq. (1) and the corrected entropy of Schwarzschild black hole. Although the constructed equivalence relation is based on a specific case, it is feasible enough for us to make some corrections for the generalized uncertainty principle effectively. To establish the equivalence relation, we will reproduce the generalized uncertainty principle from the corrected entropy. The entropy of Schwarzschild black hole corrected by the generalized uncertainty principle in Eq. (1) is provided by [25]:

$$S_G \simeq \frac{\alpha \pi k_B}{16}\left[\frac{2}{1-\sqrt{1-\frac{\alpha M_p{}^2}{16M^2}}} - \ln\left(1-\sqrt{1-\frac{\alpha M_p{}^2}{16M^2}}\right) + \ln\left(1+\sqrt{1-\frac{\alpha M_p{}^2}{16M^2}}\right)\right.$$

$$\left. + \frac{2}{\left(1+\sqrt{1-\frac{\alpha M_p{}^2}{16M^2}}\right)^2}\right], \tag{2}$$

where $\alpha$ is a positive parameter, $k_B$ is the Boltzmann constant and $M_p = \sqrt{\hbar c/G} = c^2 L_p/G$ is the Planck mass. From here onwards, the subscript $G$ is used to represent the physical quantities corrected by the generalized uncertainty principle, and the physical quantities without any subscript are regarded as the uncorrected results from the Heisenberg uncertainty principle.

We differentiate both sides of Eq. (2) to yield:

$$dS_G \simeq \frac{256\pi k_B M^3}{\alpha M_p{}^4}\left[1-\sqrt{1-\frac{\alpha M_p{}^2}{16M^2}}\right]dM. \tag{3}$$

Bekenstein-Hawking area law is given by [26]:

$$S = \frac{k_B A}{4L_p{}^2}, \tag{4}$$

which can turn Eq. (3) into:

$$dA_G \simeq \frac{32M^2}{\alpha M_p{}^2}\left[1-\sqrt{1-\frac{\alpha M_p{}^2}{16M^2}}\right]\frac{32\pi G^2 M}{c^4}dM. \tag{5}$$

According to the metric linear element of Schwarzschild black hole space-time [24], the location of the Schwarzschild black hole horizon is $r_H = 2GM/c^2$, so the black hole horizon area becomes:

$$A = 4\pi r_H{}^2 = \frac{16\pi G^2 M^2}{c^4}, \tag{6}$$

leading to:

$$dA = \frac{32\pi G^2 M}{c^4}dM, \tag{7}$$

$$\Delta A = \frac{32\pi G^2 M}{c^4} \Delta M. \tag{8}$$

Substituting Eq. (7) into Eq. (5), we can get:

$$dA_G \simeq \frac{32M^2}{\alpha M_p{}^2} \left[1 - \sqrt{1 - \frac{\alpha M_p{}^2}{16M^2}}\right] dA, \tag{9}$$

so

$$\frac{dA_G}{dA} \simeq \frac{32M^2}{\alpha M_p{}^2} \left[1 - \sqrt{1 - \frac{\alpha M_p{}^2}{16M^2}}\right]. \tag{10}$$

For black hole absorbing or radiating particle of energy, we have:

$$\Delta M \simeq c\Delta p. \tag{11}$$

Equation (8) can be updated by the above equation as:

$$\Delta A \simeq \frac{32\pi G^2 M}{c^3} \Delta p. \tag{12}$$

Considering the definition of differential [25], we can obtain another form of $dA_G/dA$ based on Eq. (12):

$$\frac{dA_G}{dA} \simeq \frac{(\Delta A_G)_{min}}{(\Delta A)_{min}} \simeq \frac{\frac{32\pi G^2 M}{c^3}(\Delta p_G)_{min}}{\frac{32\pi G^2 M}{c^3}(\Delta p)_{min}} = \frac{(\Delta p_G)_{min}}{(\Delta p)_{min}}. \tag{13}$$

Combining Eqs. (10) and (13), the corrected momentum uncertainty is gained:

$$(\Delta p_G)_{min} \simeq \frac{32M^2}{\alpha M_p{}^2} \left[1 - \sqrt{1 - \frac{\alpha M_p{}^2}{16M^2}}\right] (\Delta p)_{min}. \tag{14}$$

The Heisenberg uncertainty principle $\Delta x \Delta p \geq \hbar / 2$ gives:

$$(\Delta p)_{min} = \frac{\hbar}{2\Delta x}, \tag{15}$$

leading to:

$$(\Delta p_G)_{min} \simeq \frac{16M^2 \hbar}{\alpha M_p{}^2 \Delta x} \left[1 - \sqrt{1 - \frac{\alpha M_p{}^2}{16M^2}}\right]. \tag{16}$$

In [27], the inverse of surface gravity is treated as a sensible choice of length scale. For the Schwarzschild black hole, the surface gravity $K = 1/(2r_H)$, so we should choose:

$$\Delta x \simeq 2r_H = \frac{4GM}{c^2}, \tag{17}$$

which can turn Eq. (16) into:

$$(\Delta p_G)_{min} \simeq \frac{\Delta x \hbar}{\alpha L_p^2}\left[1 - \sqrt{1 - \frac{\alpha L_p^2}{(\Delta x)^2}}\right]. \tag{18}$$

And then we have:

$$\Delta p_G \geq (\Delta p_G)_{min} \simeq \frac{\Delta x \hbar}{\alpha L_p^2}\left[1 - \sqrt{1 - \frac{\alpha L_p^2}{(\Delta x)^2}}\right], \tag{19}$$

leading to:

$$\frac{\alpha L_p^2}{\Delta x \hbar}\Delta p_G - 1 \geq 0 \tag{20}$$

or

$$0 \geq \frac{\alpha L_p^2}{\Delta x \hbar}\Delta p_G - 1 \gtrsim -\sqrt{1 - \frac{\alpha L_p^2}{(\Delta x)^2}}. \tag{21}$$

It is obvious that Eq. (20) would become $-1 \geq 0$ as $\alpha \to 0$, so Eq. (21) should be chosen instead of Eq. (20). Let us square both sides of Eq. (21) to gain:

$$\Delta x \Delta p_G \gtrsim \frac{\hbar}{2}\left(1 + \frac{\alpha L_p^2}{\hbar^2}(\Delta p_G)^2\right), \tag{22}$$

which is a generalized uncertainty principle derived from the entropy Eq. (2). The above inequality is almost identical to the generalized uncertainty principle Eq. (1) except that "$\geq$" is replaced with "$\gtrsim$". The reason for this difference is that some approximations (such as $\Delta M \simeq c\Delta p$ and $\Delta x \simeq 2r_H$) are used in calculations and the initial equation Eq. (2) is also obtained by applying these approximations. In other words, the appearance of "$\gtrsim$" is caused by the approximations in calculations, and "$\gtrsim$" would be turned into "$\geq$" if we could improve these approximations. However, these improvements are not the focus of our research. Considering that "$\gtrsim$" does not affect the structure of the generalized uncertainty principle and it will be avoided in theory, we believe that Eq. (1) and Eq. (22) should be essentially identical, and it means that the generalized uncertainty principle is successfully reproduced from the corrected entropy. Hence, the equivalence relation between the generalized uncertainty principle and the corrected entropy should be considered to have been set up.

Our ultimate aim is to modify the generalized uncertainty principle to keep the minimum length existing under different parameter signs. Now, because the corrected entropy and the generalized uncertainty principle are equivalent, we are allowed to bring some other correction factors into the entropy so as to indirectly modify the generalized uncertainty principle without worrying about generating unreasonable modification results. Of course, these correction factors should be independent of the generalized uncertainty principle, or else the whole process would be meaningless. By choosing an appropriate correction factor, we will derive a corresponding new generalized uncertainty principle in Sect. 3 and show the actual impacts it brings in Sect. 4 and Sect. 5.

## 3. New generalized uncertainty principle from back reaction

In this section, through the path constructed by this equivalence relation in Sect. 2, we are going to modify the generalized uncertainty principle from a corrected entropy. And we hope that the modified generalized uncertainty principle can keep a fixed and unified minimum length under different parameter signs. The difference from Sect. 2 is that we will have replaced the entropy corrected by the generalized uncertainty principle with another entropy corrected by a certain correction factor. It is obvious that entropies with different correction factors would tend to lead to different forms of the generalized uncertainty principle, so it is important for us to choose an appropriate correction factor. After a series of attempts, we find the appropriate correction factor should be the back reaction [28], which will give a correction result that meets our expectation.

Due to the nonlinearity of the Einstein equations, the perturbed gravitational metric will have an effect on the evolution of the curved space-time, and the fluctuation of the space-time caused by the metric perturbation is called back reaction [28]. The one loop back reaction can be used to modify the Hawking temperature leading to a corrected surface gravity [29], so it is obvious that the black hole entropy can also be modified by it. In [23], the entropy of Schwarzschild black hole corrected by the back reaction is obtained from quantum tunneling, namely:

$$S_N \simeq \frac{4\pi k_B G^2}{L_p^2 c^4}\left[M^2 - \beta M_p^2 \ln\left(\frac{M^2}{\beta M_p^2} + 1\right)\right], \qquad (23)$$

where we write out the omitted constant $k_B$, $G$, $c$, $L_p$ and $M_p$ in [23]. Expanding the logarithm of Eq. (23), we can get a more general form: $S_N \simeq 4\pi k_B G^2/(L_p^2 c^4)\left[M^2 - 2\beta M_p^2 \ln(M) - \beta^2 M_p^4/M^2 + \beta^3 M_p^6/(2M^4) - \cdots + \text{const (independent of } M)\right]$. And it means that Eq. (23) actually contains the higher order terms of $\beta$, agreeing with most correction results [19, 24]. From here onwards, the subscript $N$ is used to represent the physical quantities corrected by the back reaction. The parameter $\beta$ is from the modified surface gravity [30, 31] and its explicit expression is provided by [29, 32]:

$$\beta = \frac{1}{360\pi}\left(-N_0 - \frac{7}{4}N_{\frac{1}{2}} + 13N_1 + \frac{233}{4}N_{\frac{3}{2}} - 212N_2\right), \tag{24}$$

where $N_s$ is the number of fields with spin 's'.

We differentiate both sides of Eq. (23) to yield:

$$dS_N \simeq \frac{4\pi k_B G^2}{L_p^2 c^4}\left(2M - 2M\frac{1}{\frac{M^2}{\beta M_p^2} + 1}\right)dM = \frac{8\pi k_B G^2 M}{L_p^2 c^4}\frac{M^2}{M^2 + \beta M_p^2}dM. \tag{25}$$

When the black hole mass $M$ is approaching the Planck scale, the entropy in Eq. (23) might vanish leading to a questionable result. However, it will not affect our following calculations because Eq. (25) shows that the correction process is based on the differential of entropy instead of the value of entropy. And the differential in Eq. (25) should keep a nonzero value as long as $M$ does not drop to zero. Equation (25) can be updated by Eqs. (4) and (7) as:

$$dA_N \simeq \frac{M^2}{M^2 + \beta M_p^2}dA, \tag{26}$$

so

$$\frac{dA_N}{dA} \simeq \frac{M^2}{M^2 + \beta M_p^2}. \tag{27}$$

Through exactly the same procedure as Eq. (13), we have:

$$\frac{dA_N}{dA} \simeq \frac{(\Delta p_N)_{min}}{(\Delta p)_{min}}. \tag{28}$$

Combining Eqs. (27) and (28), the corrected momentum uncertainty is gained:

$$(\Delta p_N)_{min} \simeq \frac{M^2}{M^2 + \beta M_p^2}(\Delta p)_{min}, \tag{29}$$

which can be rewritten by Eq. (15) as:

$$(\Delta p_N)_{min} \simeq \frac{\hbar}{2\Delta x}\frac{1}{1 + \frac{\beta M_p^2}{M^2}}. \tag{30}$$

According to Eq. (17), the above equation is changed into:

$$(\Delta p_N)_{min} \simeq \frac{\hbar}{2\Delta x}\frac{1}{1 + \frac{16\beta L_p^2}{(\Delta x)^2}}. \tag{31}$$

And then we have:

$$\Delta p_N \geq (\Delta p_N)_{min} \simeq \frac{\hbar}{2\Delta x} \frac{1}{1 + \frac{16\beta L_p^2}{(\Delta x)^2}}, \qquad (32)$$

leading to:

$$\Delta x \Delta p_N \gtrsim \frac{\hbar}{2} \frac{1}{1 + \frac{16\beta L_p^2}{(\Delta x)^2}}, \qquad (33)$$

which is a new generalized uncertainty principle derived from the entropy corrected by the back reaction Eq. (23). The above inequality is quite different from the generalized uncertainty principle Eq. (1), and their differences are mainly reflected in two aspects: on the one hand, the right-hand part of Eq. (33) is described by $(\Delta x)^2$ instead of $(\Delta p)^2$. On the other hand, $(\Delta p)^2$ in Eq. (1) appears as a second order term while $(\Delta x)^2$ in Eq. (33) appears as a complex fraction. Although Eq. (33) and Eq. (1) are formally unrelated, their parameter signs follow the same changing process. For Eq. (33), its parameter expression is given by Eq. (24) and provides a reasonable explanation for the change of its parameter sign, namely: the parameter sign depends on the spins of leading fields. For Eq. (1), a similar parameter expression is shown in [12], and the only difference from Eq. (24) is that the coefficient $1/(360\pi)$ is rewritten as $2/(45\pi)$, which does not affect the changing process of the parameter sign. In fact, the sign consistency of the two parameters ensures that the new generalized uncertainty principle Eq. (33) can directly solve the contradiction caused by the parameter sign of Eq. (1). On the contrary, if the changes of their parameter signs are inconsistent, then the cases of positive and negative parameters of Eq. (33) might not contain all the cases of Eq. (1) exactly and the subsequent results would be incomplete.

There are some details about Eq. (33) that we need to clarify: first, the sign "$\gtrsim$" will not affect the main structure of the new generalized uncertainty principle, as we discussed in Sect. 2. Second, the subscript $N$ is set up for the simplifications of calculations and itself does not limit the value range of $\Delta p_N$. Based on these considerations, we will use the following simplified inequality to replace Eq. (33) in future applications:

$$\Delta x \Delta p \geq \frac{\hbar}{2} \frac{1}{1 + \frac{16\beta L_p^2}{(\Delta x)^2}}. \qquad (34)$$

## 4. Parameter adaptability for the minimum length

In this section, we are going to show how the new generalized uncertainty principle maintains the existence of a fixed and unified minimum length under different parameter signs. In order to reflect the change of the parameter sign clearly, the absolute value of parameter is added to turn the new generalized uncertainty principle Eq. (34) into:

$$\Delta x \Delta p \geq \frac{\hbar}{2} \frac{1}{1 \mp \frac{16|\beta|L_p^2}{(\Delta x)^2}}. \tag{35}$$

In the following part, we try to get minimum lengths by using Eq. (35) respectively from positive and negative aspects.

1. **When the parameter is negative**

As $\beta < 0$, Eq. (35) should become:

$$\Delta x \Delta p \geq \frac{\hbar}{2} \frac{1}{1 - \frac{16|\beta|L_p^2}{(\Delta x)^2}}. \tag{36}$$

Because there is a natural limit: $\Delta x \Delta p \geq 0$, it is obvious that Eq. (36) would be meaningless if its right-hand part is negative. Thus, we must keep the right-hand part greater than or equal to zero. That is to say, we should ensure:

$$1 - \frac{16|\beta|L_p^2}{(\Delta x)^2} \geq 0, \tag{37}$$

leading to:

$$(\Delta x)_{min} = 4\sqrt{|\beta|}L_p, \tag{38}$$

which is the minimum length for the new generalized uncertainty principle with a negative parameter $\beta$. When the minimum length is reached, we can see the corresponding change of $\Delta p$ from Eq. (36):

$$\Delta p \geq \frac{\hbar}{2} \frac{1}{\Delta x - \frac{16|\beta|L_p^2}{\Delta x}} \to \infty, \tag{39}$$

which means the momentum uncertainty is not limited and the maximum momentum will be removed in the case of negative parameter.

2. **When the parameter is positive**

As $\beta > 0$, Eq. (35) should become:

$$\Delta x \Delta p \geq \frac{\hbar}{2} \frac{1}{1 + \frac{16|\beta|L_p^2}{(\Delta x)^2}}. \tag{40}$$

We can transform the left-hand part of the above inequality into a perfect square trinomial of $\Delta x$ and obtain:

$$\left(\Delta x - \frac{\hbar}{4\Delta p}\right)^2 \geq \frac{\hbar^2}{16(\Delta p)^2} - 16|\beta|L_p^2. \tag{41}$$

There is a natural limit: $(\Delta x - \hbar/(4\Delta p))^2 \geq 0$, so if we want to get an effective connection between $\Delta x$ and $\Delta p$ from Eq. (41), we must let the right-hand part of Eq. (41) be greater than or equal to zero, namely:

$$\frac{\hbar^2}{16(\Delta p)^2} - 16|\beta|L_p^2 \geq 0, \tag{42}$$

which immediately gives the maximum momentum:

$$(\Delta p)_{max} = \frac{\hbar}{16\sqrt{|\beta|}L_p}. \tag{43}$$

Now, we take the square root of Eq. (41) to gain:

$$\Delta x \geq \sqrt{\frac{\hbar^2}{16(\Delta p)^2} - 16|\beta|L_p^2} + \frac{\hbar}{4\Delta p}, \tag{44}$$

which provides an expression of the minimum length described by $\Delta p$:

$$(\Delta x)_{min} = \left(\sqrt{\frac{\hbar^2}{16(\Delta p)^2} - 16|\beta|L_p^2} + \frac{\hbar}{4\Delta p}\right)_{min}. \tag{45}$$

Because the part in bracket of Eq. (45) decreases as $\Delta p$ increases, the minimum value in bracket can be reached as $\Delta p = (\Delta p)_{max}$. According to Eq. (43), we have:

$$(\Delta x)_{min} = \sqrt{\frac{\hbar^2}{16(\Delta p)_{max}^2} - 16|\beta|L_p^2} + \frac{\hbar}{4(\Delta p)_{max}} = 4\sqrt{|\beta|}L_p, \tag{46}$$

which is the minimum length for the new generalized uncertainty principle with a positive parameter $\beta$. The minimum length is reached as $\Delta p = (\Delta p)_{max} = \hbar/(16\sqrt{|\beta|}L_p)$, and it means that both the minimum length and the maximum momentum can be given at the same time in the case of positive parameter.

From the above two aspects, we can see that the new generalized uncertainty principle can keep a fixed and unified value $4\sqrt{|\beta|}L_p$ as the minimum length under different parameter signs. In addition, the new generalized uncertainty principle also predicts the appearance of the maximum momentum when the parameter is positive, while the maximum momentum will be removed when the parameter is negative.

## 5. Comparisons and analyses

In order to further show the advantages of the new generalized uncertainty principle (NGUP) provided by Eq. (34), we give various generalized uncertainty principles including KMM [17], Nouicer [19], Pedram [20] and CH [21] to compare with NGUP:

$$\Delta x \Delta p \geq \frac{\hbar}{2}\left(1 + \frac{\beta L_p^2}{\hbar^2}(\Delta p)^2\right) \quad \text{(KMM)}, \tag{47}$$

$$\Delta x \Delta p \geq \frac{\hbar}{2} e^{\frac{\beta L_p^2}{\hbar^2}(\Delta p)^2} \quad \text{(Nouicer)}, \tag{48}$$

$$\Delta x \Delta p \geq \frac{\hbar}{2} \frac{1}{1 - \frac{\beta L_p^2}{\hbar^2}(\Delta p)^2} \quad \text{(Pedram)}, \tag{49}$$

$$\Delta x \Delta p \geq \frac{\hbar}{2}\left(-\frac{\beta L_p}{\hbar}\Delta p + \frac{1}{1 - \frac{\beta L_p}{\hbar}\Delta p}\right) \quad \text{(CH)}. \tag{50}$$

It should be pointed out: although the parameters of different generalized uncertainty principles might own different values or forms, now we use a common notation $\beta$ to represent them so that more intuitive comparisons can be established. Depending on Eqs. (34), (47), (48), (49) and (50), figures can be painted to reflect the differences between NGUP and other generalized uncertainty principles. From here onwards, we set $G = \hbar = c = 1$ to simplify the drawing of functions, and take $|\beta| = 1/16^2$ for NGUP and $|\beta| = 1$ for the others to keep a unified value for the maximum momentum.

In figure 1, for the case of positive parameter, all generalized uncertainty principles do not intersect the abscissa, leading to the existences of their corresponding minimum lengths. And the maximum momentum is built up for Pedram and CH to avoid the divergence of value, and for NGUP to prevent imaginary number from appearing (see Eq. (44) for more details). However, in figure 2, for the case of negative parameter, only CH and NGUP can keep their minimum lengths existing, while other generalized uncertainty principles intersect the abscissa to remove their minimum lengths. And the maximum momentum is built up for KMM to prevent $\Delta x$ from being negative (the negative value contradicts the definition of $\Delta x$). In figure 3, the minimum lengths from

CH for positive and negative parameters are obviously different because the pink dotted lines representing the position of the minimum length have separated from each other. On the contrary, in figure 4, the minimum lengths from NGUP for positive and negative parameters are exactly identical because the pink dotted lines representing the position of the minimum length have completely overlapped.

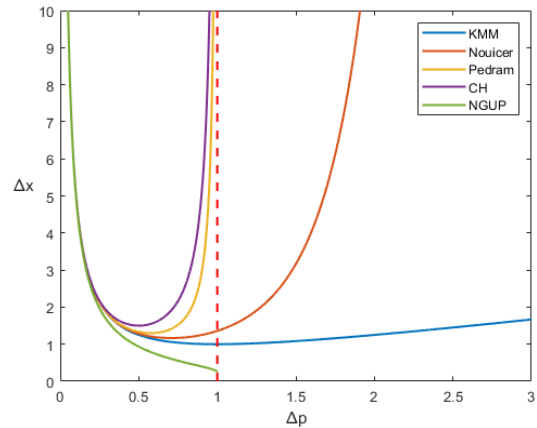

**Figure 1**. The $\Delta x$ as the functions on $\Delta p$ for $\beta > 0$ respectively given by KMM, Nouicer, Pedram, CH and NGUP. The red dotted line represents the position of the maximum momentum.

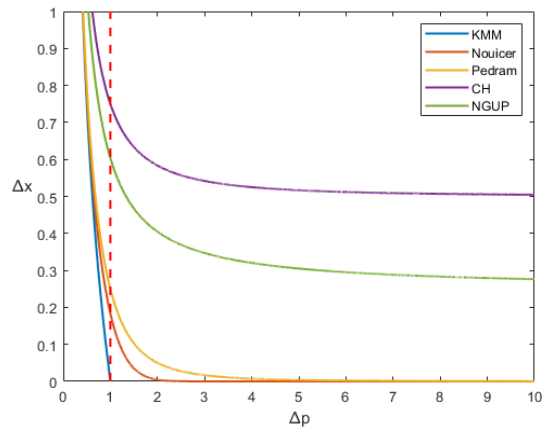

**Figure 2**. The $\Delta x$ as the functions on $\Delta p$ for $\beta < 0$ respectively given by KMM, Nouicer, Pedram, CH and NGUP. The red dotted line represents the position of the maximum momentum.

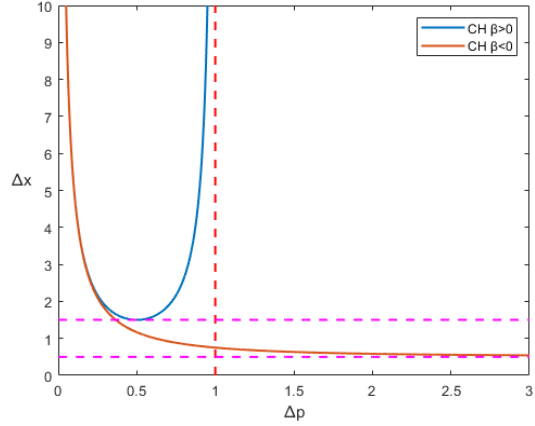

**Figure 3**. The $\Delta x$ as the functions on $\Delta p$ given by CH for $\beta > 0$ and $\beta < 0$. The red and pink dotted lines respectively represent the positions of the maximum momentum and the minimum length.

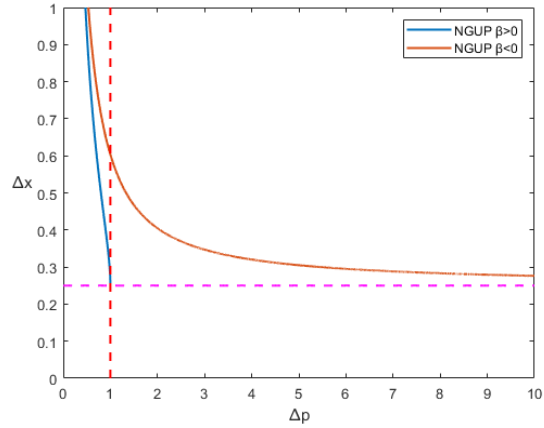

**Figure 4**. The $\Delta x$ as the functions on $\Delta p$ given by NGUP for $\beta > 0$ and $\beta < 0$. The red and pink dotted lines respectively represent the positions of the maximum momentum and the minimum length.

All the features from figure 1, figure 2, figure 3 and figure 4 can be contained in table 1 in a much clearer way. In table 1, for the minimum length, we can see that these generalized uncertainty principles including KMM, Nouicer and Pedram cannot maintain their minimum lengths in the case of negative parameter, while CH and NGUP can do that. However, the minimum lengths from CH are inconsistent in the two cases: $(\Delta x)_{min} = (3/2)|\beta|L_p$ as $\beta > 0$, but $(\Delta x)_{min} = (1/2)|\beta|L_p$ as $\beta < 0$ (these results can be obtained from Eq. (50)). By contrast, NGUP can give a fixed and unified minimum length, namely: $(\Delta x)_{min} = 4\sqrt{|\beta|}L_p$ for both $\beta > 0$ and $\beta < 0$ (the same conclusions of CH and NGUP can be drawn from figure 3 and figure 4). In theory, the minimum length should be independent of most systems and keep a stable value [13, 14], so NGUP plays a better role in reflecting the minimum length under different parameter signs than CH.

| Name | Form | $(\Delta x)_{min}$ $\beta > 0$ | $(\Delta x)_{min}$ $\beta < 0$ | $(\Delta x)_{min}$ unified | $(\Delta p)_{max}$ $\beta > 0$ | $(\Delta p)_{max}$ $\beta < 0$ |
|---|---|---|---|---|---|---|
| KMM [17] | $\Delta x \Delta p \geq \frac{\hbar}{2}\left(1 + \frac{\beta L_p^2}{\hbar^2}(\Delta p)^2\right)$ | ✓ | ✗ | ✗ | ✗ | ✓ |
| Nouicer [19] | $\Delta x \Delta p \geq \frac{\hbar}{2} e^{\frac{\beta L_p^2}{\hbar^2}(\Delta p)^2}$ | ✓ | ✗ | ✗ | ✗ | ✗ |
| Pedram [20] | $\Delta x \Delta p \geq \frac{\hbar}{2} \frac{1}{1 - \frac{\beta L_p^2}{\hbar^2}(\Delta p)^2}$ | ✓ | ✗ | ✗ | ✓ | ✗ |
| CH [21] | $\Delta x \Delta p \geq \frac{\hbar}{2}\left(-\frac{\beta L_p}{\hbar}\Delta p + \frac{1}{1 - \frac{\beta L_p}{\hbar}\Delta p}\right)$ | ✓ | ✓ | ✗ | ✓ | ✗ |
| NGUP | $\Delta x \Delta p \geq \frac{\hbar}{2} \frac{1}{1 + \frac{16 \beta L_p^2}{(\Delta x)^2}}$ | ✓ | ✓ | ✓ | ✓ | ✗ |

**Table 1**. Comparisons of various generalized uncertainty principles. "$(\Delta x)_{min}$ unified" means that the minimum lengths for $\beta > 0$ and $\beta < 0$ are unified. The tick stands for the state of including and the cross stands for the state of excluding.

In table. 1, for the maximum momentum, we can see that Pedram, CH and NGUP can lead to the maximum momentum when the parameter is positive. It means that NGUP not only has a good parameter adaptability for the minimum length but also reflects the maximum momentum supported by most high-order generalized uncertainty principles [20, 21]. In addition, only KMM provides the maximum momentum when the parameter is negative, which can restore the Chandrasekhar limit corrected by the generalized uncertainty principle [12]. It seems that the maximum momentum should also be maintained under different parameter signs, just like the minimum length. And if it is true, the generalized uncertainty principle might need to be further modified to own a parameter adaptability for both the minimum length and the maximum momentum. However, it is beyond the scope of our paper and we hope this problem can be solved in the future.

## 6. Discussions and conclusions

In fact, the calculation method that we have used in Sect. 2 and Sect. 3 is the reverse process of the improved method shown in [25]. Although there are other calculation methods [33, 34] to modify the black hole entropy by the generalized uncertainty principle, they use some obvious approximations so that the equivalence relation between the generalized uncertainty principle and the corrected entropy is too hard to be set up. For example, if we regard $A \geq B$ as $A \simeq B$, then we could not get $A \geq B$ from $A \simeq B$. By contrast, [25] gives an improved method on this point, and it suggests that we should regard $A \geq B$ as $(A)_{min} = B$. As a result, $A \geq B$ can be obtained from $(A)_{min} = B$. On the other hand, the improved method describes a more reasonable black hole evaporation [25], and it indicates that the black hole evaporation indirectly guarantees the establishment of the equivalence relation. Besides, it is worth noting that the equivalence relation in this paper depends on the Schwarzschild black hole, so another equivalence relation might need to be constructed if we want to consider electric charge, angular momentum or other factors excluded by the Schwarzschild black hole to modify the generalized uncertainty principle.

In this paper, we have derived the new generalized uncertainty principle Eq. (34) to reflect a fixed and unified minimum length in both cases of positive and negative parameters, and this is based on a natural premise: the minimum length is a model-independent feature of quantum gravity [13, 14]. However, in [35], a classical limit without the minimum length might be approached in the high energy limit and/or for large masses. Moreover, the Planck constant is even considered to be related to energy and hence seems to go to zero at high energy [36, 37], leading to the disappearance of the minimum length. Because these above scenarios are gained under special assumptions or extreme conditions, we continue to believe that the minimum length should exist in general cases. That is to say, the obtained new generalized uncertainty principle can play its role in a wide range to maintain the independence of the minimum length.

In conclusion, we first clarify an equivalence relation between the generalized uncertainty principle in Eq. (1) and the corrected entropy of Schwarzschild black hole in Eq. (2). And the equivalence relation ensures that the generalized uncertainty principle can be effectively modified by improving the entropy without causing other unnecessary modification results. After that, through the path constructed by the equivalence relation, we derive a new generalized uncertainty principle Eq. (34) from an entropy corrected by the back reaction. Comparing various generalized uncertainty principles, we can see that only the new generalized uncertainty principle can reflect a fixed and unified minimum length under different parameter signs. In other words, the new generalized uncertainty principle has a better parameter adaptability for the minimum length than other generalized uncertainty principles. In addition, the method used in Sect. 2 and Sect. 3 is universal. The method allows us put other black hole entropies corrected by different correction factors into it so as to derive corresponding generalized uncertainty principles modified by these factors. Thus, through the method, more new forms of the generalized uncertainty principle are expected to be explored.